\documentclass[prb,preprint]{revtex4-1} % style for Physical Review B and AJP are similar
\usepackage{float}     % frueher: \usepackage{here}
\usepackage{placeins}  % noetig fuer \floatbarrier
\usepackage{subfig}    %noetig fuer \subfloat
\usepackage{units}     %noetig fuer \unit[...]{...}
\usepackage{color}     %noetig fuer "highlighting etc"
\usepackage[normalem]{ulem}   %noetig fuer "\sout{bla}  %durchstreichen

\usepackage[justification=RaggedRight, singlelinecheck=false, margin=10pt,font=small,labelfont=bf]{caption} % => caption linksbuendig
\usepackage{amsmath} % need for subequations

\usepackage{graphicx} % for figures
\begin{document}

\title{A graphical description\\
of optical parametric generation of squeezed states of light}
\author{J{\"o}ran Bauchrowitz}
%%%\altaffiliation[Also at ]{home.} % optional
\email{Joeran.Bauchrowitz@aei.mpg.de} %optional
\author{Tobias Westphal}
\author{Roman Schnabel}
\affiliation{Institut f{\"u}r Gravitationsphysik, Leibniz Universit{\"a}t Hannover and Max-Planck-Institut f{\"u}r Gravitationsphysik (Albert-Einstein-Institut), Callinstra{\ss}e 38, D-30167 Hannover, Germany}

\date{\today}

\begin{abstract}
The standard process for the production of strongly squeezed states of light is optical parametric amplification (OPA) below threshold in dielectric media such as LiNbO$_3$ or periodically poled KTP. %\cite{1opa,2sq}
Here, we present a graphical description of squeezed light generation via OPA. It visualizes the interaction between the nonlinear dielectric polarization of the medium and the electromagnetic quantum field. We explicitly focus on  the transfer from the field's ground state to a squeezed vacuum state and from a coherent state to a bright squeezed state by the medium's second-order nonlinearity, respectively. Our pictures visualize the phase dependent amplification and deamplification of quantum uncertainties and give the phase relations between all propagating electro-magnetic fields as well as the internally induced dielectric polarizations. The graphical description can also be used to describe the generation of nonclassical states of light via higher-order effects of the non-linear dielectric polarization such as four-wave mixing and the optical Kerr effect.
\end{abstract}

\maketitle

\section{Introduction}

Squeezed states of light belong to the class of so-called \emph{nonclassical} quantum states. They have applications in the research field of quantum information, \cite{1qi, 2qi, 3qi, 4qi, 5qi} and were used to demonstrate quantum teleportation \cite{1qt, 2qt, 3qt} and the Einstein-Podolsky-Rosen paradox. \cite{1epr, 2epr, 3epr, 4epr} They also have applications in quantum metrology. \cite{qm} Recently, they have been applied to a  gravitational wave detector to improve its signal-to-noise ratio beyond the photon counting (shot-noise) limit. \cite{5d}
%to large scale gravitational wave detectors to improve their signal-to-noise ratio beyond the photon counting (shot-noise) limit. \cite{5d, 6d} 

\begin{figure}[h!]
%\centering
 \includegraphics[width=6 in]{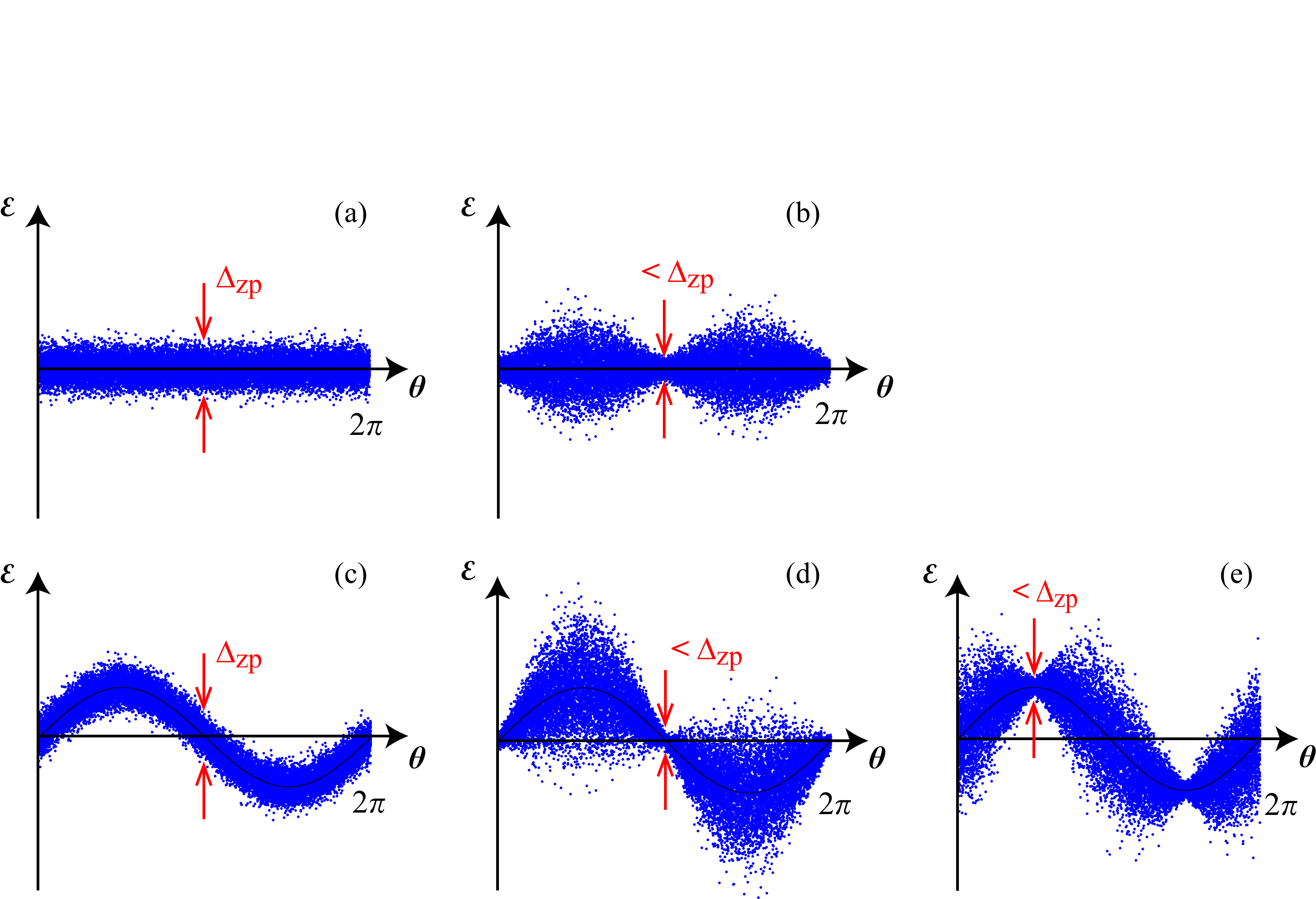}
 \caption{\cite{5s} Statistics of electric field measurements for five different minimum uncertainty states of the same optical mode. 
   (a) illustrates the ground state and its zero-point (vacuum) fluctuation $\Delta_\text{zp}$. The uncertainty does not depend on the quadrature phase $\theta$. 
   (b) represents a squeezed vacuum state. Such a state is produced by a phase dependent (optical parametric) amplification of the zero-point fluctuation, 
   (c) illustrates a  coherent state, i.e.~a displaced vacuum state, and 
   (d) and (e) are bright phase and amplitude squeezed states. For all these states the uncertainty product of the electric fields at orthogonal phases meets the lower bound set by the Heisenberg Uncertainty Relation. The above pictures are just illustrations, however, they can be experimentally reproduced by quantum state tomography using the beat signal with a homodyne local oscillator field of the same frequency.\cite{3s}
   }
\label{fig:sqz}
\end{figure}

A quantum field is said to be in a squeezed state\cite{1s, 2s, 3s, 4s} if its uncertainty is smaller than its \emph{zero-point fluctuation} for some finite range of the field's quadrature phase $\theta$ ($0 \leq \theta \leq 2\pi$). Due to the Heisenberg Uncertainty Relation, the orthogonal range of quadrature phases must then have an uncertainty larger than the zero-point fluctuation, at least by the inverse squeezing factor. Fig.~\ref{fig:sqz}a illustrates the (phase independent) zero-point fluctuation of an electric field over a full cycle of the phase from $0$ to $2 \pi$. Such a field does not have any photons on average and is said to be in its ground state (vacuum state). The quantum uncertainty of a \emph{squeezed vacuum} state is phase dependent as can be seen in Fig.~\ref{fig:sqz}b. In this work we present a graphical description that visualizes how a nonlinear dielectric medium transfers a vacuum state by help of a second harmonic field into a squeezed vacuum state. The same approach is also able to visualize the transformation from a (bright) coherent state, as shown in Fig.~\ref{fig:sqz}c to a bright phase squeezed state Fig.~\ref{fig:sqz}d or to a bright amplitude squeezed state Fig.~\ref{fig:sqz}e, respectively. 

Squeezed states of light were produced for the first time in 1985 by Slusher \textit{et al.} using four-wave mixing in a beam of Na atoms.\cite{1sq} In principle any nonlinear process, such as second-harmonic generation \cite{SHG-SQZ1, SHG-SQZ2} or the electro-optical Kerr effect, \cite{kerr1, kerr2, kerr3} can produce quadrature squeezing. The most successful process for squeezed light production is optical parametric amplification (OPA) below threshold. With this method, squeezing figures of up to $\unit[12.7]{dB}$ noise reduction have been achieved. \cite{2sq, 3sq} The same OPA process, then called parametric down-conversion,  also forms the basis for the production of entangled photon pairs\cite{1epp, 2epp}.   
OPA is based on the dielectric polarization in optical crystals with a high second-order susceptibility $\chi^{(2)}$ such as %magnesium oxide--doped
lithium niobate (%MgO:
LiNbO$_3$) or %periodically poled
potassium titanyl phosphate (%PP
KTP). % (KTiOPO$_4$)
In most squeezing experiments\cite{1opa, 2msexp, 3msexp, 3s, 5msexp, 6msexp, 7msexp, 8msexp, 9msexp} an ultra-violet or visible second harmonic \textit{pump field}  ${\cal E}(2f)$ is focused into the crystal. The pump field produces a non-linear separation of charges and thus a nonlinear dielectric polarization. The latter gives rise to a new propagating electro-magnetic field that now contains new frequency components. In this nonlinear interaction ingoing zero-point fluctuations at frequency $f$ are transformed into squeezed fluctuations. 
In this work we present an expanded graphical description of squeezed light generation by OPA. Starting from zero-point fluctuations or coherent states entering the pumped crystal, our model illustrates how the squeezed state is produced and how the different frequency components of the outgoing field are related in phase.

\section{The non-linear polarization of a dielectric medium}

We first recall the non-linear polarization of a dielectric medium. For simplicity we consider a single-path process and restrict ourselves to one spatial direction. All fields are thus scalar quantities.  The pump power is assumed to be below threshold, the noise of the pump field and the frequency dependence of the electric susceptibility $\chi(f)$ are neglected. These approximations are well justified for a description of today's squeezing experiments.

The dielectric polarization ${\cal P}$ that is caused by an electric field $\cal{E}$ inside a nonlinear medium can then be expanded in the following way  
   \begin{align}
   \label{eq:Polarisation} %Formel ueber mehrere Zeilen
      {\cal P}({\cal E}) =    \underbrace{\epsilon_0 \chi^{(1)}{\cal E}}_{{\cal P}^{(1)}}
                           +  \underbrace{\epsilon_0 \chi^{(2)}{\cal E}^2}_{{\cal P}^{(2)}}
                           +  \underbrace{\epsilon_0 \chi^{(3)}{\cal E}^3}_{{\cal P}^{(3)}}
                           + \cdots .
   \end{align}

Here,
${\cal P}^{(i)}$ is the i-th order of polarization, $\epsilon_0$ is the permittivity of vacuum and $\chi$ the electric susceptibility with typical values of $\chi^{(1)} \approx 1$, $\chi^{(2)} \approx \unit[10^{-12}]{\frac{m}{V}}$, $\chi^{(3)} \approx \unit[10^{-24}]{\frac{m^2}{V^2}}$ for state-of-the-art solid state nonlinear optical materials. \cite{1x}
For OPA-squeezing below threshold and hence for low field intensities, higher-order susceptibilities are negligibly small and the polarization can be simplified by using only the linear and the quadratic term. 

Now assume an electromagnetic field with amplitude $A$ at the optical frequency $f = \omega / 2 \pi$ that shall be squeezed, superimposed with a pump field with amplitude $B$ at twice the optical frequency inside the non-linear material. The total electromagnetic field is ${\cal E}_\text{}=A \cos(\omega t+\phi)- B \cos(2\omega t)$. After interaction with a nonlinear crystal (described by Eq.~(\ref{eq:Polarisation})), the expression for the second-order polarization of the crystal reads
   \begin{align} 
   \begin{array}{ccccc}     \label{eq:Polarisation2}
       {\cal P}^{(2)}({\cal E}_\text{})\\ 
               = \epsilon_0 \chi^{(2)} \!\!
              &  \{ A^2 \cos^2(\omega t+\phi)
              &+&   B^2 \cos^2(2\omega t)  
              &    -2AB \cos(\omega t+\phi)\cos(2\omega t)      \}  \\ 
              = \epsilon_0 \chi^{(2)} \!\!
              & \{\tfrac{1}{2}A^2[1+\underbrace{\cos(2\omega t+2\phi)}_{\propto 2\omega}]
              &+& \tfrac{1}{2}B^2[1+\underbrace{\cos(4\omega t)}_{\propto 4\omega}]   %\\
              & -AB[\underbrace{\cos(\omega t - \phi)}_{\propto \omega }
                 +  \underbrace{\cos(3\omega t + \phi)}_{\propto 3\omega }]  \}
       %+{\cal O}^3 
%      .
   \end{array}
   \end{align}
The crystal's second order polarization thus contains a DC component and components at frequencies $\omega$, $2\omega$, $3\omega$, and $4\omega$.
The component ${\cal P}^{(2)}_{\omega}=-\epsilon_0 \chi^{(2)} AB\cos(\omega t - \phi)$ interferes with the fundamental frequency component of the first-order polarization ${\cal P}^{(1)}_{\omega}=\epsilon_0 \chi^{(1)} A \cos(\omega t+\phi)$ giving rise to OPA. If all coefficients are positive, setting $\phi=\pm 90^\circ,\pm 270^\circ,...$ the fundamental input field is amplified, setting $\phi=0^\circ,\pm 180^\circ,...$ the fundamental input field is deamplified. This optical-parametric amplification process not only holds for coherent amplitudes and their classical fluctuations but also for quantum fluctuations.

%%%%%%%%%%%%%%%%%%   The graphical description   %%%%%%%%%%%%%%%%%%%%%%%%%%%%

\begin{figure}[h]%[h!]
%\centering
\includegraphics[width=6 in]{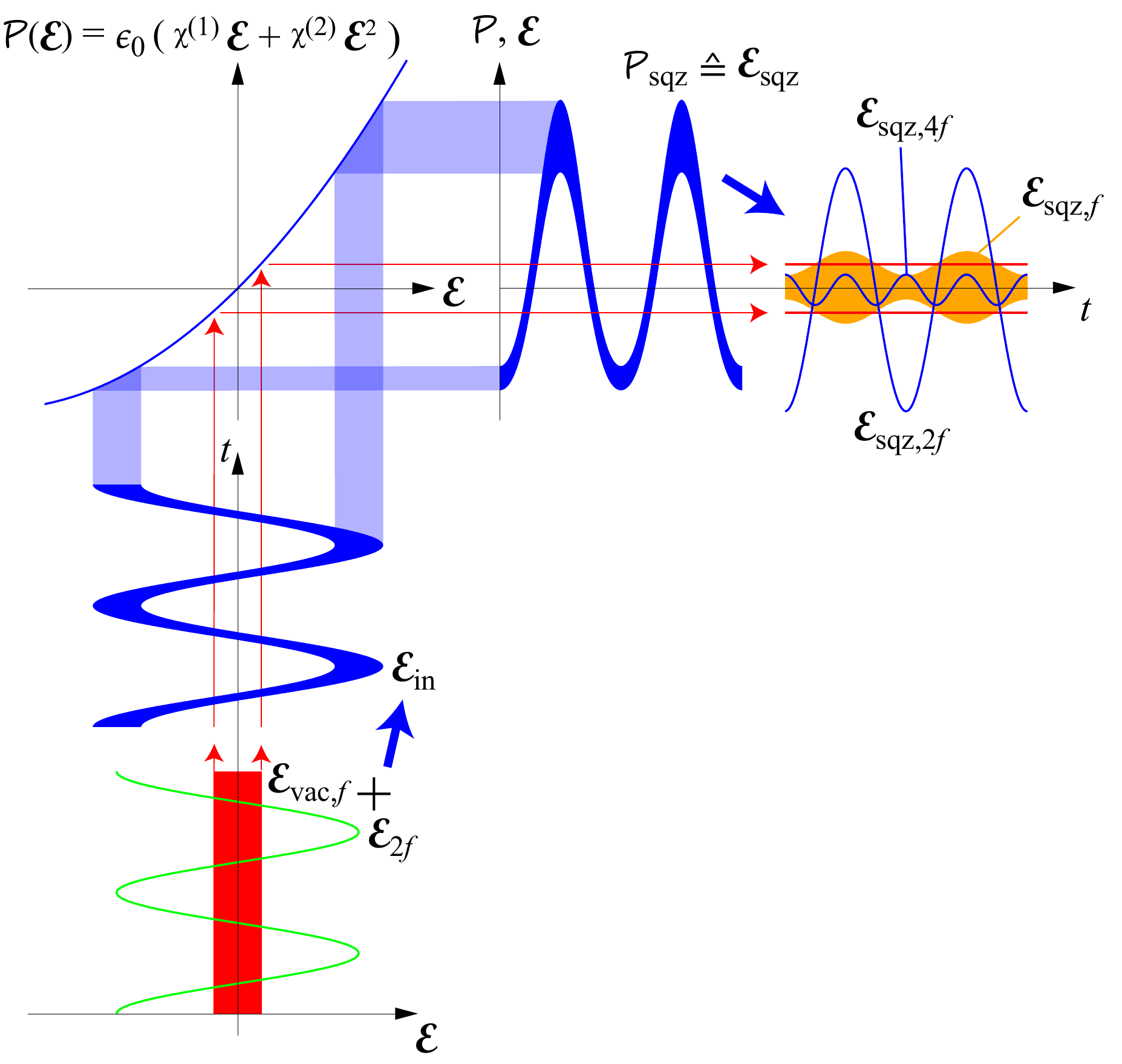}
\caption{The polarization ${\cal P}({\cal E}) =  \epsilon_0 \left(\chi^{(1)}{\cal E}  +  \chi^{(2)}{\cal E}^2\right)$ (upper left graph) describes the separation of charges of a second order non-linear material by the electric component of an optical input field. We use this graph to illustrate how an input quantum field (from below) is projected into an output quantum field (towards the right). In the example shown here, the input field is composed of a classical pump field at frequency $2f$ and zero-point fluctuations of a field at frequency $f$. The superposition of these two fields is transferred into a time-dependent dielectric polarization that is the source of (and thus directly proportional to) the electric component of the output field. The quantum uncertainty of the output field shows a phase dependent amplification with frequency $2f$. Spectral decomposition of the output field ${\cal E}_\text{sqz}$ reveals coherent amplitudes at frequencies $2f$ and $4f$ and a squeezed vacuum state ${\cal E}_{\text{sqz,}f}$ as shown in Fig.~\ref{fig:sqz}b.
}     
\label{fig:Polarisation-eines-Medium_sqz_ampl_vac__simpel}
\end{figure}

\section{The graphical description of optical parametric generation of squeezed states of light}

Our graphical description combines the usual way of displaying quantum fields \cite{3s} with that one illustrating the effect of the dielectric polarization inside a medium in terms of a ${\cal P}({\cal E})$-diagram.\cite{2x} Uncertainties of time-domain quantum fields are usually represented as areas along the time axis with a width that corresponds to the standard deviation of the uncertainty. If an electric input field uncertainty is projected by the ${\cal P}({\cal E})$-diagram from the ${\cal E}$-axis to the ${\cal P}$-axis the transfer to the uncertainty of the dielectric polarization becomes obvious. Since the latter is proportional to the radiated output field, the overall nonlinear transfer of quantum noise due to the nonlinear dielectric polarization is depicted. 
Our first example is given in Fig.~\ref{fig:Polarisation-eines-Medium_sqz_ampl_vac__simpel} and describes the conversion of a vacuum state into a squeezed vacuum state via OPA. All input fields enter the graph from below. The relevant electric field components are the zero-point fluctuations at the fundamental frequency ${\cal E}_{\text{vac,}f}$ and a classical pump field at the harmonic frequency ${\cal E}_{2f}$.  The total field causes a nonlinear separation of charges ${\cal P}_\text{sqz}$ inside the crystal. 
The graph shows that the interplay between the two fields results in a phase-dependent amplification and deamplification of the quantum uncertainty at the fundamental frequency. Apart from the quantum noise ${\cal E}_{\text{sqz,}f}$, classical fields at frequencies $2f$ and $4f$ leave the dielectric medium. The amplitude at frequency $2f$ is connected to the pump field's first-order polarization ${\cal P}^{(1)}({\cal E}_{2f})$ and the amplitude at frequency $4f$ is connected to its second-order polarization ${\cal P}^{(2)}({\cal E}_{2f})$. The comparison of the outgoing quantum noise ${\cal E}_{\text{sqz,}f}$ with the ingoing vacuum field ${\cal E}_{\text{vac,}f}$ (horizontal lines across ${\cal E}_{\text{sqz,}f}$) reveals the squeezing effect. 
The presence of the second harmonic pump field ${\cal E}_{2f}$ is obviously crucial to produce strong squeezing since it harmonically drives the input uncertainty along the characteristic curve. The pump field maxima produce an amplification of the uncertainty, minima lead to deamplification of the uncertainty. Both happens twice per fundamental period. The stronger the pump field the stronger the parametric amplification. Small uncertainties of the pump field do not play a significant role in this process. Vacuum input fields at other frequencies than $f$ are not converted into a stationary squeezed vacuum state.\\

\begin{figure}[h!]%[h!]
%\centering
    \includegraphics[width=6 in]{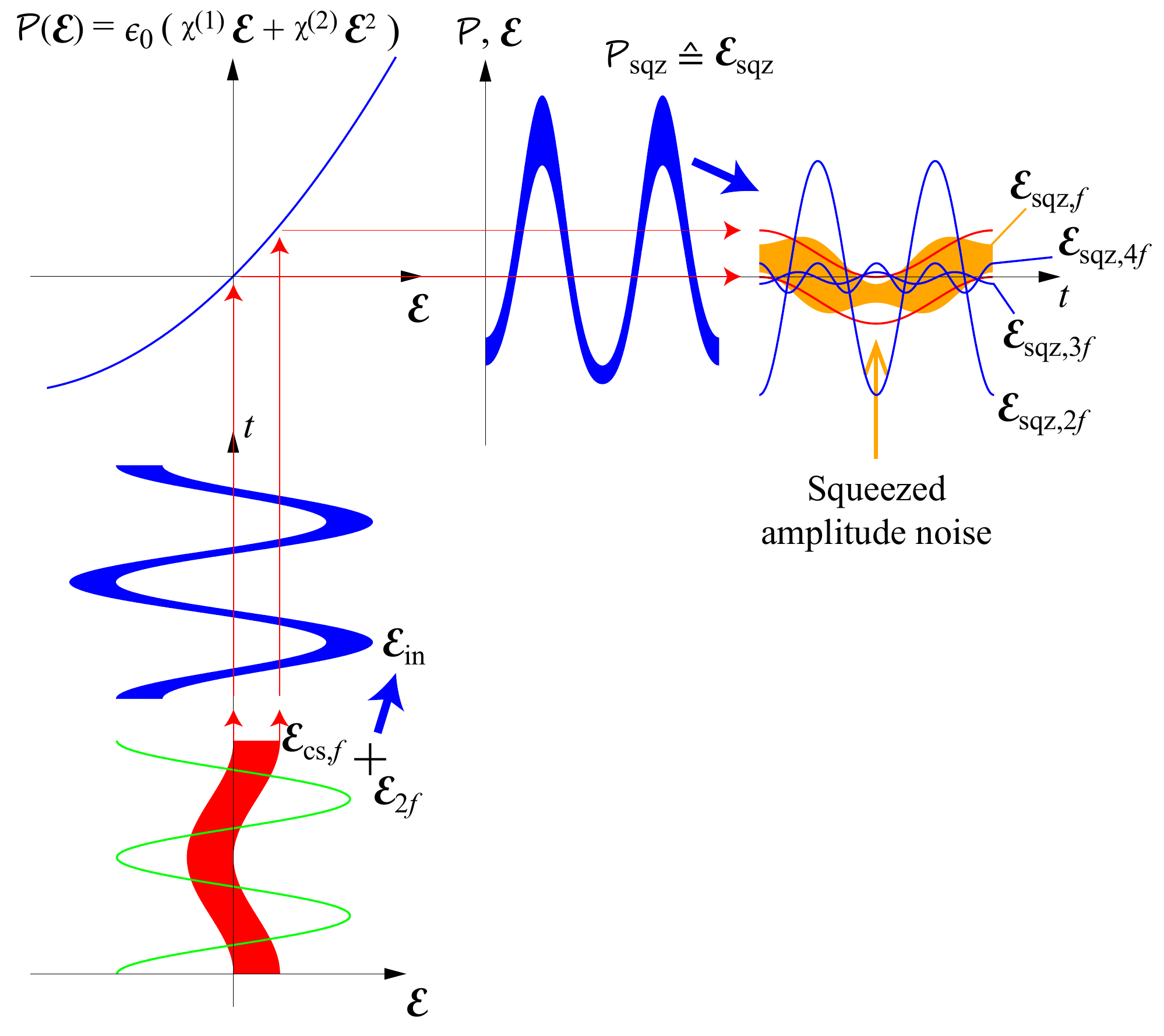}
\caption{In this example the input field ${\cal E}_{\text{in}}$ consists of a displaced vacuum state ${\cal E}_{\text{cs,}f}$ (coherent state) at frequency $f$ (cf. Fig.~\ref{fig:sqz}c) and a classical pump field amplitude at $2f$. The phase between the two components is chosen such that minima of the second harmonic field coincide with extrema of the fundamental field. The overall output field is again shown in blue. It is composed of classical amplitudes at frequencies $2f$, $3f$ and $4f$ (cf. Eq.~(\ref{eq:Polarisation2})) as well as an amplitude squeezed state at the fundamental frequency (cf. Fig.~\ref{fig:sqz}e). It is obvious from the figure that a phase shift of the harmonic input field by $180^\circ$ would result in a phase squeezed output field as shown in Fig.~\ref{fig:sqz}d.
 }
\label{fig:Polarisation-eines-Medium_sqz_ampl_power}
\end{figure}

\FloatBarrier  % noetig: \usepackage{placeins}
                        % Alle Bilder, die oberhalb dieses Befehls im Quelltext stehen werden oberhalb positioniert.

The second example is given in Fig.~\ref{fig:Polarisation-eines-Medium_sqz_ampl_power}. Here, a coherent state ${\cal E}_{\text{cs,}f}$ together with its second harmonic pump field ${\cal E}_{\text{}2f}$ enter the picture from below. Their relative phase determines what type of squeezed state is produced. For the phase chosen in Fig.~\ref{fig:Polarisation-eines-Medium_sqz_ampl_power} the coherent displacement at the fundamental frequency is deamplified, and so is the uncertainty of the field's amplitude. The uncertainty area of the input field is thus converted into the depicted uncertainty of the output field as shown on the right side of the figure. It belongs to an amplitude squeezed state as shown in Fig.~\ref{fig:sqz}e. Squeezed states having a coherent displacement are sometimes called \textit{bright} squeezed states.\cite{bright} 
The output field also has higher-order frequency components at $2f$, $3f$ and $4f$ that need to be separated to extract the state at fundamental frequency $f$. In experimental squeezed light generation the higher order frequencies are suppressed by destructive interference, e.g. filtered out by dichroic beam splitters. 
Phase shifting the second harmonic pump field by half of its wavelength results in an amplified coherent displacement at the fundamental frequency exhibiting phase squeezing.

\section{CONCLUSIONS}
We have presented a graphical picture that describes the conversion of vacuum states and coherent states of light to squeezed states via optical parametric amplification. It combines the quantum uncertainties of optical fields with the nonlinear dielectric polarization of the crystal medium. The latter's uncertainty as induced by the input field is also included. Our picture does, however, not explain the general origin of quantum uncertainties. Those are quantified by the Heisenberg Uncertainty Relation and are taken here for granted.
In our graphical description the phase dependent amplification of the electric field is introduced as a purely classical process. This is in full agreement with quantum physics, as long as electric field strengths are considered. Nevertheless, \emph{squeezing} is a sufficient condition of nonclassicality. If we move from electric field measurements to photon counting the nonclassicality of squeezed states becomes obvious since they show a sub-poissonian counting statistics. Our picture thus does not present a 'classical' description of squeezed states, which is generally not possible, but offers a physical description how squeezed states are generated in state-of-the-art experiments.
Our graphical approach can be expanded in a straight forward manner to describe the effect of higher order polarizations on quantum uncertainties such as four-wave mixing and the Kerr effect, \cite{kerr1, kerr2, kerr3}.

\begin{acknowledgments}
We acknowledge support from the Centre for Quantum Engineering and Space-Time Research (QUEST), the International Max Planck Research School on Gravitational
Wave Astronomy (IMPRS). We also thank Klemens Hammerer, Sebastian Steinlechner and Alexander Khalaidovski for helpful discussions.% and ... for proofreading.}

\end{acknowledgments}


\begin{thebibliography}{5}

%%%%%%%%%%%%%%%%%%%%%%%%%%%%%%%%
%%%   MATH: squeezing        %%%
%%%%%%%%%%%%%%%%%%%%%%%%%%%%%%%%

%\bibitem{1d}D. W. Robinson,
%``The Ground State of the Bose Gas,'' Commun. math. Phys. {\bf 1}, 159--174
%(1965).

%\bibitem{2d}D. Stoler,
%``Equivalence Classes of Minimum Uncertainty Packets,'' Phys. rev, D {\bf 1}, 3217
%(1970).

%\bibitem{3d}D. Stoler,
%``Equivalence Classes of Minimum-Uncertainty Packets. II,'' Phys. rev. D {\bf 4}, 1925--1926
%(1971).

%\bibitem{4d}E. Y. C. Lu,
%``New Coherent States of the Electromagnetic Field,'' Lettere Al Nuovo Cimento Series 2 {\bf 2}, Nr. 24, 1241--1244
%(1971).



%%%%%%%%%%%%%%%%%%%%%
%%%   Unschärfe   %%%
%%%%%%%%%%%%%%%%%%%%%

%\bibitem{u}W. Heisenberg,
%``{\"U}ber den anschaulichen Inhalt der quantentheoretischen Kinematik und Mechanik,'' Zeitschrift f{\"u}r Physik A Hadrons and Nuclei {\bf 43}, Nr.3, 172--198
%(1927).






%%%%%%%%%%%%%%%%%%%%%%%%%%%%%%%%%%%%%%%%%%
%%%   quantum information processing   %%%
%%%%%%%%%%%%%%%%%%%%%%%%%%%%%%%%%%%%%%%%%%


\bibitem{1qi}H. P. Yuen and J. H. Shapiro,
``Optical Communication with Two-Photon Coherent States -- Part I: Quantum-State Propagation and Quantum-Noise Reduction,'' IEEE Trans. Inf. Theory {\bf 24}, Nr. 6, 657--668
(1978).

\bibitem{2qi}Y. Yamamoto and H. A. Haus,
``Preparation, measurement and information capacity of optical quantum states,'' Rev. Mod. Phys. {\bf 58}, 1001--1020
(1986).

\bibitem{3qi}B. E. A. Saleh and M. C. Teich,
``Can the channel capacity of a light-wave communication system be increased by the use of photon-number--squeezed light?,'' Phys. Rev. Lett. {\bf 58}, 2656--2659
(1987).

\bibitem{4qi}S. L. Braunstein and P. v. Loock,
``Quantum information with continuous variables,'' Rev. Mod. Phys. {\bf 77}, 513--577
(2005).

\bibitem{5qi}G. Smith, J. A. Smolin, and J. Yard,
``Quantum communication with Gaussian channels of zero quantum capacity,'' Nat. Phot. {\bf 5}, 624--627
(2011).

%%%%%%%%%%%%%%%%%%%%%%%%%%%%%%%%%
%%%   quantum teleportation   %%%
%%%%%%%%%%%%%%%%%%%%%%%%%%%%%%%%%

\bibitem{1qt}A. Furusawa, J. L. S{\o}rensen, S. L. Braunstein, C. A. Fuchs, H. J. Kimble, and E. S. Polzik,
``Unconditional quantum teleportation,'' Science {\bf 282}, 706--709
(1998).

\bibitem{2qt}W. P. Bowen, N. Treps, B. C. Buchler, R. Schnabel, T. C. Ralph, H.-A. Bachor, T. Symul, and P. K. Lam,
``Experimental investigation of continuous-variable quantum teleportation,'' Phys. Rev. A {\bf 67}, 032302
(2003).

\bibitem{3qt}J. DiGuglielmo, B. Hage, A. Franzen, J. Fiur{\'a}{\v{s}}ek, and R. Schnabel,
``Experimental characterization of Gaussian quantum-communication channels,'' Phys. Rev. A {\bf 76}, 012323
(2007).




%%%%%%%%%%%%%%%%%%%%%%%
%%%   EPR paradow   %%%
%%%%%%%%%%%%%%%%%%%%%%%

\bibitem{1epr}A. Einstein, B. Podolsky, and N. Rosen,
``Can {Quantum-Mechanical} Description of Physical Reality Be Considered Complete?,'' Phys. Rev. {\bf 47}, 777--780
(1935).

\bibitem{2epr}Z. Y. Ou, S. F. Pereira, H. J. Kimble, and K. C. Peng,
``Realization of the Einstein-Podolsky-Rosen Paradox for Continuous Variables,'' Phys. Rev. Lett. {\bf 68}, Nr. 25, 3663--3666
(1992).

\bibitem{3epr}W. P. Bowen, R. Schnabel, and P. K. Lam,
``Experimental Investigation of Criteria for Continuous Variable Entanglement,'' Phys. Rev. Lett. {\bf 90}, Nr. 4, 043601
(2003).

\bibitem{4epr}S. Steinlechner, J. Bauchrowitz, T. Eberle, and R. Schnabel,
``Strong EPR-steering with unconditional entangled states,'' 
%Preprint at http://arxiv.org/abs/1112.0461
(2011). arXiv:1112.0461.


%%%%%%%%%%%%%%%%%%%%%%%%%%%%%
%%%   quantum metrology   %%%
%%%%%%%%%%%%%%%%%%%%%%%%%%%%%

\bibitem{qm}R. Schnabel, N. Mavalvala, D. E. McClelland, and P. K. Lam,
``Quantum metrology for gravitational wave astronomy,'' Nat. Comm. {\bf 1}, Nr. 121, 1--10
(2010).




%%%%%%%%%%%%%%%%%%%%%%%%%%%%%%%%%%%%%%%%%%%%%%%
%%%   SQZ-Anwendung: Theorie/squeezed GEO   %%%
%%%%%%%%%%%%%%%%%%%%%%%%%%%%%%%%%%%%%%%%%%%%%%%


%%%% Sascha würden nicht dieses Paper:
%\bibitem{5d}H. Vahlbruch,
%``The GEO 600 squeezed light source,'' Class. Quantum Grav. {\bf 27}, 084027
%(2010).
%%%% ...sondern:
\bibitem{5d}The {LIGO} Scientific Collaboration,
``A gravitational wave observatory operating beyond the quantum shot-noise limit,'' Nature Physics {\bf 7}, Nr. 12, 962--965
(2011).


%\bibitem{6d}{LIGO} Scientific Collaboration,
%``Enhancing the astrophysical reach of the LIGO gravitational wave detector by using
%squeezed states of light,'' (to be published).






%%%%%%%%%%%%%%%%%%%%
%%%   squeezed   %%%
%%%%%%%%%%%%%%%%%%%%


\bibitem{1s}H. P. Yuen,
``Two-photon coherent states of the radiation field,'' Phys. Rev. A {\bf 13}, 2226--2243
(1976).

\bibitem{2s}D. F. Walls,
``Squeezed states of light,'' Nature {\bf 306}, 141--146
(1983).

\bibitem{3s}G. Breitenbach, S. Schiller, and J. Mlynek,
``Measurement of the quantum states of squeezed light,'' Nature {\bf 387}, 471-475
(1997).

\bibitem{4s}V. V. Dodonov,
``{`Nonclassical'} states in quantum optics: a `squeezed' review of the first 75 years,'' J. Opt. B Quantum Semiclassical Opt. {\bf 4}, R1--R33
(2002).


\bibitem{5s}C. M. Caves,
``Quantum-mechanical noise in an interferometer,'' Phys. Rev. D {\bf 23}, Nr. 8, 1693--1708
(1981).








%%%%%%%%%%%%%%%%%%%%%%%%%%%%%%%
%%%   Erster Sqz, SHG-Sqz   %%%
%%%%%%%%%%%%%%%%%%%%%%%%%%%%%%%


\bibitem{1sq}R. E. Slusher, L. W. Hollberg, B. Yurke, J. C. Mertz, and J. F. Valley,
``Observation of Squeezed States Generated by Four-Wave Mixing in an Optical Cavity,'' Phys. Rev. Lett. {\bf 55}, Nr. 22, 2409--2412
(1985).


%\bibitem{SHG-SQZ}L. A. Lugiato,
%``Squeezed states in second-harmonic generation,'' Opt. Lett. {\bf 8}, Nr. 5, 256--258
%(1983).


% Erstes SHG-SQZ-Experiment:
\bibitem{SHG-SQZ1}S. F. Pereira, M. Xiao, H. J. Kimble, and J. L. Hall,
``Generation of squeezed light by intracavity frequency doubling,'' Phys. Rev. A {\bf 38}, Nr. 9, 4931--4934
(1988).

% Stärkstes SHG-SQZ-Experiment (2.4dB):
\bibitem{SHG-SQZ2}H. Tsuchida,
``Generation of amplitude-squeezed light at 431 nm from a singly resonant frequency doubler,'' Opt. Lett. {\bf 20}, Nr. 21, 2240--2242
(1995).


%%%%%%%%%%%%%%%%
%%%   Kerr   %%%
%%%%%%%%%%%%%%%%

\bibitem{kerr1}K. Bergman and H. A. Haus,
``Squeezing in fibers with optical pulses,'' Opt. Lett. {\bf 16}, Nr. 9, 663--665
(1991).

\bibitem{kerr2}K. S. Zhang, T. Coudreau, M. Martinelli, A. Ma{\^i}tre, and C. Fabre, %Echtes Kerr-Experiment
``Generation of bright squeezed light at \unit[1.06]{$\mu$m} using cascaded nonlinearities in a triply resonant cw periodically-poled lithium niobate optical parametric oscillator,'' Phys. Rev. A {\bf 64}, 033815
(2001).

\bibitem{kerr3}R. Dong, J. Heersink, J. F. Corney, P. D. Drummond, U. L. Andersen, and G. Leuchs,
``Experimental evidence for Raman-induced limits to efficient squeezing in optical fibers,'' Opt. Lett. {\bf 33}, Nr. 2, 116--118
(2008).



%\bibitem{kerr?}M. Kitagawa,
%``Number-phase minimum-uncertainty state with reduced number uncertainty in a Kerr nonlinear interferometer,'' Phys. Rev. A {\bf 34}, Nr. 5, 3974--3988
%(1986).

%\bibitem{kerr?}A. G. White,
%``Kerr noise reduction and squeezing,'' J. Opt. B.: Quantum Semiclass. Opt. {\bf 2}, 553--561
%(2000).




%%%%%%%%%%%%%%%%%%%%%%%%%%%%%%%%%%%%%%
%%%   10 dB   OPA-squeezer   %%%
%%%%%%%%%%%%%%%%%%%%%%%%%%%%%%%%%%%%%%


\bibitem{2sq}H. Vahlbruch, M. Mehmet, S. Chelkowski, B. Hage, A. Franzen, N. Lastzka, S. Go{\ss}ler, K. Danzmann, and R. Schnabel,
``Observation of Squeezed Light with 10-dB Quantum-Noise Reduction,'' Phys. Rev. Lett. {\bf 100}, 033602
(2008).



%%%%%%%%%%%%%%%%%%%%%%%%%%%%%%%%%%%%%%
%%%   bester OPA-squeezer   %%%
%%%%%%%%%%%%%%%%%%%%%%%%%%%%%%%%%%%%%%

\bibitem{3sq}T. Eberle, S. Steinlechner, J. Bauchrowitz, V. H{\"a}ndchen, H. Vahlbruch, M. Mehment, H. M\"{u}ller-Ebhardt, and R. Schnabel,
``Quantum Enhancement of the Zero-Area Sagnac Interferometer Topology for Gravitational Wave Detection,'' Phys. Rev. Lett. {\bf 104}, Nr. 25, 251102
(2010)





%%%%%%%%%%%%%%%%%%%%%%
%%%   Type I/II    %%%
%%%%%%%%%%%%%%%%%%%%%%


%\bibitem{3sq}J. E. Midwinter,
%``The effects of phase matching method and of uniaxial
%crystal symmetry on the polar distribution of second-order non-linear optical polarization,'' Brit. J. Appl. Phys. {\bf 16}, Nr. 8, 1135--1142
%(1965).






%%%%%%%%%%%%%%%%%%%%%%%%%%%%%%%%%%
%%%   entangled photon pairs   %%%
%%%%%%%%%%%%%%%%%%%%%%%%%%%%%%%%%%

\bibitem{1epp}C. K. Hong, Z. Y. Ou, and L. Mandel,
``Measurement of Subpicosecond Time Intervals between Two Photons by Interference,'' Phys. Rev. Lett. {\bf 59}, Nr. 18, 2044--2046
(1987).

\bibitem{2epp}D. Bouwmeester, J.-W. Pan, K. Mattle, M. Eibl, H. Weinfurter, and A. Zeilinger,
``Experimental quantum teleportation,'' Nature {\bf 390}, 575--579
(1997).



%%%%%%%%%%%%%%%%%%%%%%%%%%%%%%%%%%%%%%
%%%   erster OPA-squeezer   %%%
%%%%%%%%%%%%%%%%%%%%%%%%%%%%%%%%%%%%%%

\bibitem{1opa}L.-A. Wu, H. J. Kimble, J. L. Hall, and H. Wu,
``Generation of Squeezed States by Parametric Down Conversion,'' Phys. Rev. Lett. {\bf 57}, Nr. 20, 2520--2523
(1986).




%%%%%%%%%%%%%%%%%%%%%%%%%%%%%%%%
%%%   most sqz experiments   %%%
%%%%%%%%%%%%%%%%%%%%%%%%%%%%%%%%

\bibitem{2msexp}A. Ourjoumtsev, R. Tualle-Brouri, J. Laurat, P. Grangier,
``Generating Optical Schr\"{o}dinger Kittens for Quantum Information Processing,'' Science {\bf 312}, 83--86
(2006).

\bibitem{3msexp}J. Laurat, T. Coudreau, N. Treps, A. Ma{\^i}tre, and C. Fabre,
``Conditional Preparation of a Quantum State in the Continuous Variable Regime: Generation of a sub-Poissonian State from Twin Beams,'' Phys. Rev. Lett. {\bf 91}, Nr. 21, 213601
(2003).

\bibitem{5msexp}S. Feng and O. Pfister,
``Quantum Interference of Ultrastable Twin Optical Beams,'' Phys. Rev. Lett. {\bf 92}, Nr. 20, 203601
(2004).

\bibitem{6msexp}A. S. Villar, L. S. Cruz, K. N. Cassemiro, M. Martinelli, and P. Nussenzveig,
``Generation of Bright Two-Color Continuous Variable Entanglement,'' Phys. Rev. Lett. {\bf 95}, 243603
(2005).

\bibitem{7msexp}Y. Takeno, M. Yukawa, H. Yonezawa, and A. Furusawa,
``Observation of \unit[-9]{dB} quadrature squeezing with improvement of phase stability in homodyne measurement,'' Opt. Expr. {\bf 15}, Nr. 7, 4321--4327
(2007).

\bibitem{8msexp}J. S. Neergaard-Nielsen, B. M. Nielsen, C. Hettich, K. M{\o}lmer, and E. S. Polzik,
``Generation of a Superposition of Odd Photon Number States for Quantum Information Networks,'' Phys. Rev. Lett. {\bf 97}, 083604
(2006).

\bibitem{9msexp}P. K. Lam, T. C. Ralph, B. C. Buchler, D. E. McClelland, H.-A. Bachor, and J. Gao,
``Optimization and transfer of vacuum squeezing from an optical parametric oscillator,'' J. Opt. B {\bf 1}, 469-474
(1999).

%%%%%%%%%%%%%%%%%%%%%%%%%%%%%%%%%
%%%   Susze   %%%
%%%%%%%%%%%%%%%%%%%%%%%%%%%%%%%%%

%\bibitem{1x}L. Bergmann,
%``Lehrbuch der Experimentalphysik: Band 3,'' 10. Auflage, Gruyter
%(2004).

\bibitem{1x}Robert W. Boyd,
\textsl{Nonlinear Optics}, 3rd ed. 
(Academic Press, Amsterdam [et al.], 2008), pp. 1--3.

\bibitem{2x}Bahaa E. A. Saleh and Malvin C. Teich,
\textsl{Fundamentals of photonics}, 1st ed.
(Wiley, New York [et al.], 1991), pp. 739--746.

\bibitem{bright}F. A. M. Oliveira and P. L. Knight,
``Bright Squeezing,'' Phys. Rev. Lett. {\bf 61}, Nr. 7, 830--833
(1988).



\end{thebibliography}
\end{document}